\documentstyle[prl,aps,epsf]{revtex}

\def\cO{{\cal O}}
 
\def\gtwid{\raise.3ex\hbox{$>$\kern-.75em\lower1ex\hbox{$\sim$}}}
\def\ltwid{\raise.3ex\hbox{$<$\kern-.75em\lower1ex\hbox{$\sim$}}}

\def\eg{{\it e.g.},\ }
\def\et{{\it et al.}}
\def\prl#1{Phys.\ Rev.\ Lett.\ {\bf #1}}
\def\prd#1{Phys.\ Rev.\ {\bf D#1}}
\def\plb#1{Phys.\ Lett.\ {\bf #1B}}
\def\npb#1{Nucl.\ Phys.\ {\bf B#1}}

\def\MeV{{\rm Me\!V}}
\def\GeV{{\rm Ge\!V}}
\newcommand{\fB}{$f_B$}
\newcommand{\fBs}{$f_{B_s}$}
\newcommand{\fD}{$f_D$}
\newcommand{\fDs}{$f_{D_s}$}
\newcommand{\fBsofB}{$f_{B_s}/f_B$}
\newcommand{\fDsofD}{$f_{D_s}/f_D$}
\newcommand{\fBofDs}{$f_B/f_{D_s}$}

\begin{document}
\twocolumn[\hsize\textwidth\columnwidth\hsize\csname
@twocolumnfalse\endcsname

\draft
\title{
Lattice Determination of Heavy-Light Decay Constants
}
\author{
C.~Bernard$^1$, 
T.\ DeGrand$^2$,
C.\ DeTar$^3$,
Steven Gottlieb$^4$, Urs M.~Heller$^5$, J.\ E.~Hetrick$^6$,
N.~Ishizuka$^7$, C.\ McNeile$^3$,
R.~Sugar$^8$, D.~Toussaint$^9$,
M.~Wingate$^{10}$
}
\address{
$^1$Washington University, St.~Louis, Missouri 63130, USA\\
$^2$University of Colorado, Boulder, Colorado 80309, USA \\
$^3$University of Utah, Salt Lake City, Utah 84112, USA\\
$^4$Indiana University, Bloomington, Indiana 47405, USA\\
$^5$SCRI, The
Florida State University, Tallahassee,
   Florida 32306-4130, USA\\
$^6$University of the Pacific, Stockton, CA 95211-0197, USA\\
$^7$University of Tsukuba, Tsukuba Ibaraki 305, Japan\\ 
$^8$University of California, Santa Barbara, California 93106, USA\\
$^9$University of Arizona, Tucson, Arizona 85721, USA\\
$^{10}$RIKEN BNL Research Center, Upton, New York 11973, USA
}
\date{\today}

\maketitle


\begin{abstract}\noindent
We report on the MILC collaboration's calculation of \fB, \fBs, \fD, 
\fDs\  and their ratios.  
Our central values
come from the quenched approximation, but 
the quenching error is estimated from $N_F\!=\!2$ 
dynamical staggered lattices.
We 
use Wilson light valence quarks 
and Wilson and static heavy quarks.  
We find, for example,
$f_B=157 \pm 11\ {}^{+25}_{-9}\  {}^{+23}_{ -0}\  \MeV$,
$f_{B_s}/f_B  = 1.11 \pm 0.02\ {}^{+0.04}_{ -0.03}    \pm 0.03$,
$f_{D_s}  = 210 \pm 9\ {}^{+25}_{ -9} \ {}^{+17}_{ -1}\  \MeV$ and 
$f_{B}/f_{D_s}  = 0.75 \pm 0.03\ {}^{+0.04}_{ -0.02}\ {}^{+0.08}_{ -0.00}  $,
where the errors are statistical, systematic (within the quenched
approximation), and systematic (of quenching), respectively.
\end{abstract}
\pacs{PACS numbers: 12.15.Hh, 12.38.Gc, 13.20.-v}
]

The neutral B meson ($B_d$), a bound state of a $d$ quark and an
anti-$b$ quark, is known to mix with its antiparticle,
${\overline B}_d$.
In the Standard Model, $x_d$, the ratio of the mixing oscillation to the
decay rate, is proportional to the absolute square
of the
fundamental quantity
$V_{td}$.  However, despite the fact that the
$x_d$ is well measured \cite{BBbarEXPT}, $V_{td}$
remains poorly determined because the proportionality constant
between $x_d$ and $|V_{td}|^2$ depends on nonperturbative strong
interaction  effects.  These effects  are parameterized by 
$f_B$,
the pseudoscalar decay constant of the $B_d$ meson, and $B_B$, 
the corresponding ``bag parameter.''  Accurate computations of $f_B$ and $B_B$
therefore put tight constraints on the Standard Model.
Similarly, a measurement of
$x_s$ for $B_s$ mesons  would determine
a second fundamental quantity,
$V_{ts}$, if $f_{B_s}$ and $B_{B_s}$ were known, or $|V_{td}/V_{ts}|$,
if the ratios $f_{B_s}/f_B$ and $B_{B_s}/B_B$ were known.

Lattice QCD offers a way to compute quantities like $f_B$ and $B_B$
from first principles.  Here, we present a 
computation by the MILC collaboration of the decay constants
\fB, \fBs, \fD, \fDs, and their ratios.
Ref.~\cite{LAT98} gives additional details;
preliminary results were described in 
Refs.~\cite{PRELIM-MILC-SETUP,PRELIM-MILC}.

Table~\ref{tab:lattices} shows the lattice parameters used.  
See \cite{PRELIM-MILC-SETUP} for details of the lattice generation, 
gauge fixing, and determination of the
quark propagators.
We compute ``smeared-local'' and ``smeared-smeared'' pseudoscalar
meson propagators in each of three cases: heavy-light, static-light,
and light-light (with degenerate masses only).  
Light Wilson quark propagators are computed by a minimal residual
algorithm
for three values of the hopping parameter, giving light quark
masses ($m_q$) in the range $0.7 m_s \ltwid m_q \ltwid 2.0 m_s$,
where $m_s$ is the strange quark mass.  
The light-light
pseudoscalars are used to set the scale (through $f_\pi$) and to
find the physical values of $\kappa_{ud}$ and $\kappa_s$, the
hopping parameters
of the up/down and strange quarks.
We determine $\kappa_s$ by adjusting the degenerate pseudoscalar mass to 
$\sqrt{2m_K^2 - m_\pi^2}$,
the lowest order chiral perturbation theory value.
We also compute 
smeared-local light-light vector meson propagators, which we
use for alternative determinations of the scale (through $m_\rho$)
and $\kappa_s$ (through $m_\phi$). 

Heavy quark propagators are computed by the hopping parameter
expansion \cite{HENTY}.
Because of practical limitations to this
approach \cite{PRELIM-MILC-SETUP}, we sum the sink
point of the smeared-smeared correlators 
only over a subset of points in a spatial volume.
This means that intermediate states of nonzero 3-momenta can contribute.
For the heavy-light mesons studied here, these
higher  momentum states are suppressed  sufficiently at asymptotic
Euclidean time $t$  by their higher energy, although in the largest volumes
(sets  N and O) this can require
$t_{\rm min}/a$ as large as 25 ($a$ is the lattice spacing).

The static-light mesons have no such suppression.  However on our
smallest volumes 
(sets A, C, D, F, G, H)
the contamination by higher momentum states  is small
($\approx 0.7\%$, which we estimate using static-light wavefunctions from
Ref.~\cite{WAVEFUNCTIONS}).
On all other sets the contamination is  expected
to be large.  We therefore have performed a dedicated static-light
computation on those lattices, with relative smearing functions taken 
from \cite{KENTUCKY} and zero momentum intermediate states enforced
by a complete FFT sum over spatial slices.  In addition, the dedicated
static light computation has been run on set A (because the plateaus from
the hopping method proved to be poor) and set G (as a check of the
hopping method).  On the latter set, the two methods give consistent
results.

For all pseudoscalars, we fit the 
smeared-local and smeared-smeared correlators 
simultaneously and covariantly to single exponential forms,
with the same mass in both channels.
We vary the fit range (in $t$) in each channel over
several choices that have reasonable confidence level (CL). 
Combining
such choices for the light-light, heavy-light and static-light
cases, we have approximately 25 different
versions of the analysis on each data set.  Our central values are taken
from the version which has the best blend of high CL 
and small statistical errors.  We then find the standard deviation of the
result over the other versions and add it in quadrature with the raw
jackknife error of the central value. 
The resulting error will be called, henceforth, 
``the statistical error.''

We employ the EKM norm \cite{EKM}
throughout. In the heavy-light case
we also adjust the measured meson
pole mass upward by the difference between 
the heavy quark kinematic mass ($m_2$) and
the heavy quark
pole mass ($m_1$) 
as calculated  in the tadpole-improved tree approximation \cite{EKM}, 
fixing the mean link from $\kappa_c$.
We use the
one-loop tadpole-improved,
mass-dependent perturbative renormalization 
of the axial current \cite{KURAMASHI}, 
with coupling $\alpha_V(3.4018/a)$ defined 
in terms of the plaquette
\cite{ALPHAV}.
We adjust the result of \cite{KURAMASHI} for our matching point
($m_2$ rather than $m_1$) and for 
our choice of the mean link.
Our central values use ``scale choice {\it i}\/'':
$q^*_{HL}=2.32/a$ for the heavy-light corrections \cite{BGM} and
$q^*_{SL}=2.18/a$ for the static-light corrections \cite{HERNANDEZandHILL}.
The heavy-light scale was calculated in the massless limit; however,
since it differs little from the static-light scale, it seems reasonable
to use it for all mass values.  The effects of two other choices of 
scale ({\it ii}: $q^*_{HL}=q^*_{SL}=1/a$; 
{\it iii}: $q^*_{HL}= 4.63/a$, $q^*_{SL}=4.36/a$) 
give an estimate of the perturbative errors.

In our chiral fits we use $am_2$ as the independent variable rather
than the more standard $1/\kappa$. 
Although the two are formally equivalent at this order in $a$,
$m_2$ has the advantage that
CL
of linear fits to 
$M_{Qq}$ and $f_{Qq}$  are quite good. (Here, $Q$ is a generic --- possibly 
static --- heavy quark, and $q$
is a generic light quark.)  Further, linear fits to $f_\pi$ also have 
reasonable 
CL
 for quenched $\beta\ge6.0$.
For $m_\pi^2$, however,
the 
CL
 of the linear fits is uniformly very poor 
whether $1/\kappa$ or $am_2$ is the independent variable.  To study this
problem in more detail, we have examined the pseudoscalar mesons for
6 light quark masses at $\beta=5.7$ on additional lattices 
(set ``5.7-large'') of size 
$12^3\times48$
(403 configurations), $16^3\times48$ (390 confs.),
$20^3\times48$ (200 confs.), and $24^3\times48$ (184 confs.).
The lightest two meson masses in this set ($\approx 385$ and 
$\approx 515\ \MeV$), are below those used in the full computation.
On set 5.7-large, linear fits of $m_\pi^2$ {\it vs.}\ either
$1/\kappa$  or $am_2$ 
are still poor, but quadratic 
fits are good.
Indeed, a quadratic fit of $m_\pi^2$ {\it vs.}\ $am_2$ using the 5 heaviest
masses goes right through the lightest $m_\pi^2$ in all but the $12^3$ volume.
For our central values we thus employ
quadratic fits {\it vs.}\ $am_2$ for $m_\pi^2$, and linear fits 
{\it vs.}\ $am_2$ for $f_\pi$, $M_{Qq}$, and $f_{Qq}$.  We call
this ``chiral fit {\it I}.''
Three other fit choices 
({\it II}: all linear; {\it III}: $m_\pi^2$ and $f_\pi$ quadratic, all
others linear; {\it IV}: $m_\pi^2$, $f_\pi$ and $f_{Qq}$ quadratic, all
others linear)
are used to assess the systematic error.   One  set, F, undergoes very large
($\sim\!50\%$) variation when the chiral fit choice is changed, 
possibly because of finite size effects.
Set F is therefore dropped from further analysis.

To find $f_B$ on a given data set, 
we divide out
the perturbative logarithms \cite{KURAMASHI}
from $f_{Qq}\sqrt{M_{Qq}}$, fit to a 
polynomial in $1/M_{Qq}$, interpolate to $m_B$, and then replace the
logarithms. 
We do three versions of the polynomial fit: (1) a quadratic fit to
the mesons in the approximate mass range $2$ 
to $4$ GeV (``heavier heavies'') (2) a quadratic fit to
the mesons in the approximate mass range $1.25$ to 
$2$ GeV (``lighter heavies'') (3) a cubic fit to
the mesons in the approximate mass range $1.25$ to $4$ GeV.  
We include the static-light point in all three fits.
We use range (1) in central values for $f_B$ and $f_{B_s}$;
range (2), for $f_D$ and $f_{D_s}$. The alternative ranges go into
the systematic error estimates.

\begin{figure}[thb]
\epsfxsize=0.99 \hsize
\epsffile{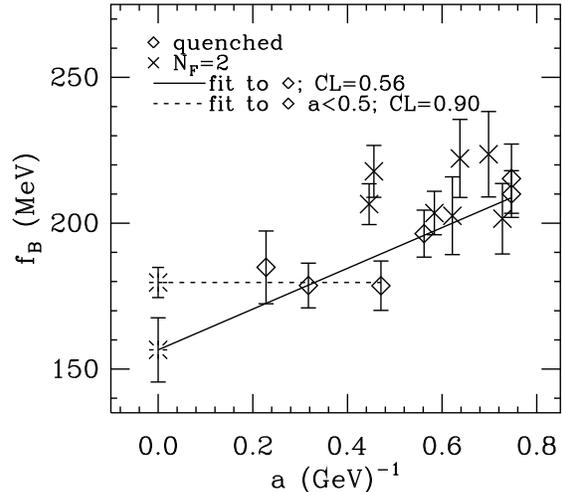}
\caption{$f_B$ {\it vs.}\ $a$ for preferred choices: chiral fit
{\it I}, perturbative scale {\it i}, and ``heavier heavies'' mass range.
The scale is set by $f_\pi$.
The linear fit to all quenched points (solid line)
gives the central value.  }
\label{fig:fbvsa}
\end{figure}

The final extrapolation is in lattice spacing. Since the Wilson action's
leading errors are $\cO(a)$, we attempt a linear extrapolation in $a$
for all our quenched results. Figure \ref{fig:fbvsa} shows the extrapolation
for $f_B$, with the central choices 
of the perturbative scale (choice {\it i})
and of the chiral fits (fit {\it I}).
An alternative possibility, with which
the data are also consistent, is that the $\cO(a)$ effects are small
enough for $6/g^2 \ge 6.0$ ($a\ltwid0.5\ \GeV^{-1}$) that one may extrapolate
with a constant fit  in this region.  
For the decay constants,
both fits have 
acceptable 
CL, 
but the constant fit is 
better.  However, for 
\fBsofB\ and \fDsofD, the  linear fits (${\rm CL}\approx0.6$)
are much better than
the constant fits (${\rm CL}\approx0.1$). (See Ref.~\cite{LAT98} for plots
of the ratios and additional details.)  Since it would
be inconsistent to treat the decay constants as independent
of $a$, yet fit the ratios linearly, and since we in any case expect
significant $\cO(a)$ errors for Wilson fermions,
we take the linear fits to the quenched results for our central values.
The differences with  the constant fits
are included in the systematic errors.  
At this point,
the dynamical $N_F=2$ data is not good enough to extrapolate 
to the continuum, even for unphysically large dynamical
quark mass.
We use the dynamical
data only to assess the error due to \hbox{quenching}.

\begin{figure}[thb]
\epsfxsize=0.99 \hsize
\epsffile{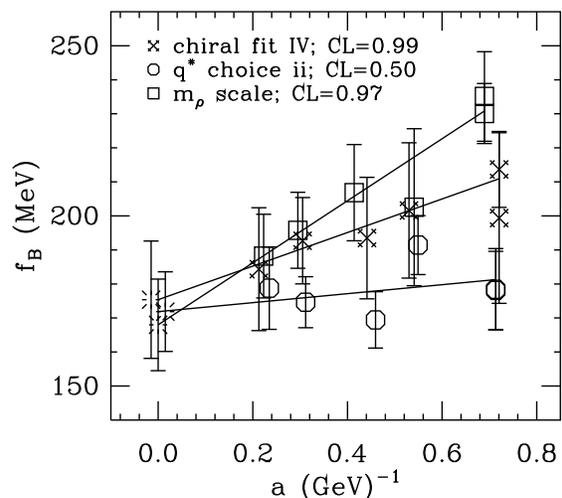}
\caption{Same as Fig.~\ref{fig:fbvsa}, but quenched results only,
with alternative analysis choices. Some points have been displaced
slightly horizontally for clarity.}
\label{fig:fbvsa_alt}
\end{figure}

The systematic errors are computed as follows:

(1) The three largest sources of error within the
quenched approximation are the continuum extrapolation,
the chiral extrapolation, and the $\cO(\alpha_s^2)$
perturbative corrections (as estimated from  a change in scale in the
$\cO(\alpha_s)$ terms).  With our data, these errors cannot be
computed independently.
For example, when the chiral extrapolations are changed to
fit {\it IV} (see Fig.~\ref{fig:fbvsa_alt}), the difference between the 
linear and constant (not shown)
continuum extrapolations gets smaller ($15\ \MeV$ instead
of $23\ \MeV$).
Further, while the 
systematic error in the final results would be
very small 
if the only source of uncertainty
were the next higher order perturbative correction ($\cO(\alpha_s^2)$),
this is not the case once the interplay between perturbative uncertainties
and other continuum extrapolation errors is included.
Indeed, assume there exists ``perfect"
data which are linear in $a$ with slope and intercept as in
Fig.~\ref{fig:fbvsa}, and then add on an $\cO(\alpha_s^2)$ correction
with  a coefficient chosen to give the same change at $\beta=5.7$ as
would be produced by reducing $q^*_{HL}$ and $q^*_{SL}$
to $1/a$ (choice {\it ii}).  Although this gives
a $17\%$ change at $\beta=5.7$, a linear extrapolation of the changes
to $a=0$ results in a residual error of less than 1\%.  With the real
data, however, changing  $q^\star$ to  choice {\it ii} raises the linearly
extrapolated value by $10\%$ (see Fig.~\ref{fig:fbvsa_alt}); while it
reduces the constant
fit by $3\%$. Choice {\it iii}\/ reduces the linear
value and raises the constant value by $\sim\!2\%$.
We therefore estimate the errors from the continuum extrapolation,
chiral extrapolation and perturbative corrections together.
We compute each quantity 24 times (2 continuum extrapolations $\times$
4 chiral fits $\times$ 3 scale choices), giving a central
value and  23 alternatives.  The alternatives
are divided
into two groups depending on whether the result is greater or less than the
central value, and the standard deviation of each group about the central
value is then taken as the positive or negative combined error.

(2) The ``magnetic mass'' $m_3$, which
divides the chromomagnetic interaction in the effective non-relativistic
Hamiltonian for Wilson fermions, is not equal 
to the kinetic mass $m_2$ \cite{EKM}.  
This introduces an error at fixed $a$
of $\cO((c_{mag}-1)\Lambda_{QCD}/M_{Qq})$,
where $c_{mag}\equiv m_2/m_3$.
The error is not completely removed by the linear extrapolation
to $a=0$.
Following \cite{JLQCD}, we estimate the residual error
by using the tree level expression for $c_{mag}$ (with our values
of $am_2$) and extrapolating $c_{mag}$ linearly
in $a$.  With our preferred
choices for the mass range in the $f_{Qq}\sqrt{M_{Qq}}$ fit,
this gives an error of $\sim\!2\%$ for $f_B$ and
$\sim\!3\%$ for $f_D$.
The error on $f_B$ can be reduced to less than $1\%$
by switching to the ``lighter heavies'' ($+$ static) mass range:
the static-light point, for which $m_3 \not= m_2$ is not
an issue,  becomes particularly
important in this case.
In practice, we assess
the errors due to $m_3 \not= m_2$ as the larger of: (a) the $2$ or $3\%$
model estimate with our preferred mass ranges  and (b) the actual difference 
in the final result caused by switching from ``heavier heavies'' to
``lighter heavies'' or {\it vice versa}.
 
(3) Our preferred fits of $f_{Qq}\sqrt{M_{Qq}}$ {\it vs}. $1/M_{Qq}$ are
truncated at quadratic order.
A scale of $\sim\!0.75$ GeV for $1/M_{Qq}$ is expected in the 
omitted cubic term,
since this is roughly the scale size found in the linear and quadratic terms.  
We calculate that the existence of such a cubic term in the data
would lead to an error, in the analysis that uses only quadratic fits,
of $\sim\!1\%$ in the decay constants.
In practice we estimate this error by
changing to cubic fits 
(using the entire mass range $1.25$
to $4$ GeV); the errors found are indeed $\ltwid1\%$.

(4) The finite volume effects are estimated by comparing results on
sets A (spatial size $\sim\!1.2$~fm) and B ($\sim\!2.5$~fm) and applying 
the fractional difference to the final results. 
Set A is smaller than all other quenched lattices; 
B, much larger.  Therefore the difference
should give a conservative bound on the finite volume error.  In practice,
we take the larger of: (a) the difference when all quantities are computed
individually on sets A and B and (b) the difference when all light-light
quantities 
are taken from set 5.7-large.
Since there is some
cancellation of error between $f_{Qq}$ and $f_\pi$, (b) is generally
larger.  We find an
error of $\sim\!2$--$3\%$ on  decay constants, $\sim\!4\%$ on  \fBofDs, and
$\sim\!1$--$2\%$ on  other ratios.

Errors (2)--(4) do not have definite signs and appear to be largely 
independent of each other and of error (1).
We thus take the error within the quenched approximation to be the sum, 
in quadrature,  of errors (1) through (4).  For decay constants,
error (1) always dominates; while for the ratios, error (2) 
(and for \fBofDs, (4)) is (are) comparable to (1).

(5) The quenching error is estimated in three ways:  (a) We 
set the scale by using $m_\rho$ instead of $f_\pi$.  (b) For
quantities involving the strange quark, we fix $\kappa_s$ from
$m_\phi$ instead of the pseudoscalars. (c) We compare the results from
the weakest coupling $N_F=2$ lattices (sets G and R) with the quenched
results interpolated ({\it via} the linear fit)
to the same value of the lattice spacing
($a\approx 0.45\ {\rm GeV}^{-1}$ --- see Fig.~\ref{fig:fbvsa}).  We average
this difference 
over 12 analysis choices (4 chiral fits $\times$ 3 scale choices), 
plus (for strange quark quantities) the
preferred choices but with $\kappa_s$ fixed from $m_\phi$.  For
the decay constants, this difference has a definite sign over all 
12 or 13 choices. We then take the signed error (c) to be just
the average difference. 
However, for some of the ratios, the standard deviation of the
difference is larger than the average difference.
In that case, the positive (negative) error (c) is taken to be 
average difference plus (minus) the standard deviation.  Finally,
the quenching error in the positive or negative
direction is defined to be the largest of errors (a), (b),
and (c) in that direction.  In almost all cases, 
(c) is largest.

Note that our quenching error estimate is still rather crude.
Our $N_F=2$ simulations are not ``full QCD'' because they are not
extrapolated to the continuum or to the physical quark mass, and they
do not have a dynamical strange quark.
For these reasons we prefer to quote the central values
as the quenched results and to treat the 
difference (c) as a signed error, not a correction. 
(See \cite{LAT98} for further discussion.)
Additional dynamical simulations, which we hope will allow us
to improve this situation, are in progress.
Note, however,
that the sign we find of the difference 
is what is expected from intuitive arguments
about the wave function at the origin
\cite{WEINGARTEN}.

We then have:

\begin{eqnarray}
&f_B = 157\ \pm 11\ {}^{+25}_{-9} \ {}^{+23}_{ -0}\ \ \MeV\cr
&f_{B_s} = 171\ \pm 10\ {}^{+34}_{ -9} \ {}^{+27}_{ -2}\ \ \MeV\cr
&f_D = 192\ \pm 11\ {}^{+16}_{ -8} \ {}^{+15}_{ -0}\ \ \MeV\cr
&f_{D_s} = 210\ \pm 9\ {}^{+25}_{ -9} \ {}^{+17}_{ -1}\ \ \MeV\cr
&f_{B_s}/f_B = 1.11\ \pm 0.02\ {}^{+0.04}_{ -0.03} \pm 0.03\cr
&f_{D_s}/f_D = 1.10\ \pm 0.02\ {}^{+0.04}_{ -0.02} \ {}^{+0.02}_{ -0.03}\cr
&f_{B}/f_{D_s} = 0.75\ \pm 0.03\ {}^{+0.04}_{ -0.02} \ {}^{+0.08}_{ -0.00}\cr
&f_{B_s}/f_{D_s} = 0.85\ \pm 0.03\ {}^{+0.05}_{ -0.03} \ {}^{+0.05}_{-0.00}\ ,
\end{eqnarray}
where the errors are statistical, systematic (within the quenched
approximation), and systematic (of quenching), respectively.
The result for $f_{D_s}$ is consistent with the experimental value
\cite{RICHMAN} of $ 241 \pm  21 \pm 30\ \MeV$.
Our quenched approximation values are consistent with 
recent quenched results using improved actions \cite{JLQCD,FERMILAB}.

This work was supported by the U.S. DOE and NSF.
Calculations for this project were performed 
at ORNL CCS, SDSC,
Indiana University,  NCSA,
PSC, MHPCC, CTC,
CHPC (Utah), 
and Sandia Natl.\ Lab.
We thank Y.\ Kuramashi for discussions,
and the Columbia group and the HEMCGC collaboration for
lattice sets F and G,
respectively.

\vspace{-0.5truein}

\vfill\eject

\begin{table}
\caption{Lattice parameters.  Sets F, G, and L--R use
variable-mass Wilson valence quarks and
two flavors of fixed-mass staggered
dynamical fermions;
all other runs use quenched
Wilson quarks.
}
\label{tab:lattices}
\begin{center}
\begin{tabular}{clrr} \hline
set& $\beta\  (am_q)$ &size &\# confs. \\
\hline
\vrule height 10pt width 0pt A  & 5.7 & $8^3\times 48$ & 200 \\
\hline
\vrule height 10pt width 0pt B  & 5.7 & $16^3\times 48$ & 100 \\
\hline
\vrule height 10pt width 0pt E& $ 5.85$&  $12^3 \times 48$&  100 \\
\hline
\vrule height 10pt width 0pt C  & 6.0 & $16^3\times 48$ & 100\\
\hline
\vrule height 10pt width 0pt D  & 6.3 & $24^3\times 80$ & 100\\
\hline
\vrule height 10pt width 0pt H& $ 6.52$&  $ 32^3 \times 100$& 60 \\
\hline
\vrule height 10pt width 0pt L& $ 5.445\ (0.025)$&  $ 16^3 \times 48$&  100 \\
\hline
\vrule height 10pt width 0pt N& $ 5.5\ (0.1)$&  $ 24^3 \times 64$&   101 \\
\hline
\vrule height 10pt width 0pt O& $ 5.5\ (0.05)$&  $ 24^3 \times 64$&   100 \\
\hline
\vrule height 10pt width 0pt M& $ 5.5\ (0.025)$&  $ 20^3 \times 64$&   199 \\
\hline
\vrule height 10pt width 0pt P& $ 5.5\ (0.0125)$&  $ 20^3 \times 64$&   199\\
\hline
\vrule height 10pt width 0pt G& $ 5.6\ (0.01)$&   $16^3 \times 32$&   200 \\
\hline
\vrule height 10pt width 0pt R& $ 5.6\ (0.01)$&   $24^3 \times 64$&    200 \\
\hline
\vrule height 10pt width 0pt F& $ 5.7\ (0.01)$&   $16^3 \times 32$&    49 
\end{tabular}
\end{center}
\end{table}


\begin{references}
\bibitem{WEINGARTEN} See, \eg
F.\ Butler \et, \npb{421}, 217 (1994). Note that although
we are really computing $f_B/f_\pi$, one would expect $f_B$
to be more affected by the potential at the origin than $f_\pi$,
because the $B$ is a smaller state.
\end{references}
\end{document}